\documentclass[useAMS,usenatbib]{mn2e}

\usepackage{graphicx}
\usepackage{amsmath}
\title[Michel H\'enon's contributions to collisional stellar systems]{Michel H\'enon's contributions to collisional stellar systems}
\author[D.C. Heggie]{Douglas C.
Heggie$^{1}$
\\
$^{1}$University of Edinburgh, School of Mathematics and Maxwell
Institute for Mathematical Sciences, King's
Buildings,\\ Edinburgh EH9 3JZ, UK; d.c.heggie@ed.ac.uk
}

\def\textcolor#1{}

\def\be{{\mathbf e}}

\def\bv{{\mathbf v}}
\def\bV{{\mathbf V}}
\def\henon{H\'enon}
\def\henons{H\'enon's}
\def\gtorder{\mathrel{\raise.3ex\hbox{$>$}\mkern-14mu
             \lower0.6ex\hbox{$\sim$}}}
\def\ltorder{\mathrel{\raise.3ex\hbox{$<$}\mkern-14mu
             \lower0.6ex\hbox{$\sim$}}}

\begin{document}
\pagerange{\pageref{firstpage}--\pageref{lastpage}} \pubyear{2002}

\maketitle


\label{firstpage}

\begin{abstract}
  The theory of star cluster dynamics was a major topic in \henons\ 
  early research career.  Here we summarise his contributions under
  three headings: (i) the Monte Carlo method, (ii) homological
  evolution of star clusters, and (iii) escape from star clusters.  In
  each case we also trace some aspects of how \henons\  contributions
  have been developed or applied in subsequent decades up to the
  present.  We also propose that \henons\  work be
  commemorated by adopting the names ``\henon\ units'' and ``\henons\ 
  Principle''. 
\end{abstract}

\section{Introduction}

\henon's contributions to collisional stellar dynamics are found in
less than 20 papers\footnote{We have used the bibliography of ADS in
  this paper}.  All were published before 1976, but represent almost
half of the output in that early period of his research career.
Though \henon\  abandoned the field thereafter, his contributions were
taken up again in subsequent decades by others, sometimes after a long
gap.  Now they have become established as essential tools in the
current vigorous exploration of the dynamical evolution of globular
star clusters.

In this contribution we summarise \henons\  research under three
headings, not in chronological order.  The last in chronological order
was the invention of the Monte Carlo method for the numerical
simulation of the evolution, and we discuss this first.  It also
involves a couple of ideas which we recommend be used to commemorate
\henons\  contributions: ``\henon\  units'' and
``\henons\  Principle''.  
Then we turn to what was essentially \henons\  thesis work,
on self-similar evolution.  Besides the self-similar models
themselves, it is a seminal work, foreshadowing so much that was
painfully rediscovered or refined by others in subsequent decades.
Finally we examine a topic to which \henon\  returned from time to time: the escape rate
from a star cluster, especially an isolated one.  This was a natural
conclusion to his earliest collisional research, on the evaluation of diffusion
coefficients, but that is the subject of another contribution in this
volume. 

\section{The Monte Carlo code (1967-1975)}

This was a subject which occupied \henon\  for about ten years \citep{
1967NASSP.153..295H,1967nmds.conf...91H,1971Ap&SS..13..284H,1971Ap&SS..14..151H,1972ASSL...31...44H,1972ASSL...31..406H,1973dses.conf..183H,1974A&A....37..183A,1975IAUS...69..133H}\footnote{The
  two
  papers in Astrophysics and Space Science are reprints  of those in
  the Proceedings of IAU Colloquium No.10}.  
In fact Monte Carlo codes were  invented independently by H\'enon, on
the  one hand, and by
  Spitzer and his students on the other, in the years around 1970, and their history is well
 summarised in \citet{V2014}.
After \henon\  himself apparently stopped developing the method in the
mid-1970s, there was a gap before it was taken up again and further
developed  by J. Stod\'o\l kiewicz in the 1980s.   Sadly Stod\'o\l kiewicz
died in 1988, but the ideas underlying his and \henons\  codes were
faithfully developed from the 1990s by Stod\'o\l kiewicz's student M. Giersz, working at
CAMK in Warsaw.  The same thread of ideas was also adopted by a group
at Northwestern University under the leadership of F. Rasio.  This
research programme has been pursued actively by both the US and the
Polish group ever since, and now have reached a quite comparable level
of development.  A version of Giersz's code, though by no means the
latest, can be found in the AMUSE software repository at
amusecode.org, where it is referred to as {\sl mmc}.

\subsection{Description of the basic Monte Carlo method}

Once \henon\  had thought of it,  the basic idea
of these Monte Carlo codes is quite simple.  Assuming that 
 a stellar system is spherical and non-rotating,  the 
 dynamics of each star is represented by its (specific) energy $E$ and
 angular momentum $J$.   The values of $E$ and $J$ are given random
 adjustments with the same statistical properties as those given  by the
 theory of two-body relaxation (including moments $\langle\Delta E\rangle, \langle(\Delta E)^2\rangle,$ etc.)

In slightly greater detail, the structure of the code is the following.
    \begin{enumerate}
\item[1] Select the overall time step $dt$, which for a code of \henon\ 
  type is a fraction of a relaxation time;
\item[2] From the initial model (King, Plummer, etc), for each star assign 
radius $r$ (distance from the cluster centre) and $E$ and $J$.
\item[3] Begin:
  \begin{enumerate}
  \item[I]
  Order the stars by $r$ and compute the potential
\item[II] For each successive pair  of stars
  \begin{enumerate}
\item From $E,J$ compute radial and transverse components of velocity of both stars.  (The orientation of the transverse component is  randomised)
\item Choose the impact parameter $p$ for an encounter so that the statistical effect of the encounter corresponds to the velocity changes predicted by the theory of relaxation for the time $dt$
\item Compute the velocity components after the encounter, and  the new $E$ and $J$ of both stars
\item Move each star to a random position (radius) on an orbit of energy $E$ and angular momentum $J$
  \end{enumerate}
  \end{enumerate}
\item[IV] Repeat 
    \end{enumerate}
Really, anyone with a knowledge of collisional stellar dynamics could
use this outline to construct a simple Monte Carlo code in a few
hours, except for a few tricky points, where \henons\  ingenuity shows
the way.  The shortest introduction to his approach is in
\citet{1972ASSL...31..406H,1971Ap&SS..14..151H}, which are the same
paper, or his SAAS-Fee lectures \citep{1973dses.conf..183H}, which
also include a succinct account of the theory of two-body
relaxation.  Here we summarise \henons\  solutions to these awkward points.

\begin{enumerate}
\item
{\sl Computation of the potential}

The first is simple enough.  In a spherical system the  potential gradient is
$\displaystyle{\frac{d\phi}{dr} = -\frac{GM(r)}{r^2}}$, where $M(r)$
is the mass inside radius $r$.  Therefore the  potential is 
  \begin{eqnarray*}
\phi &=& \int_r^\infty \frac{GM(r)}{r^2}dr    \\
&=& \left[-\frac{GM(r)}{r}\right]_r^\infty + \int_r^\infty \frac{G dM(r)}{r}.
  \end{eqnarray*}
 In the Monte Carlo code, each star is thought of as a spherical shell. 
Thus let $m_i, r_i, \phi_i$ be the mass, radius and potential of the
$i$th star or shell.  Then $\displaystyle{\phi_i =
  \sum_{j=1}^i\frac{Gm_j}{r_i} + \sum_{j=i+1}^N\frac{Gm_j}{r_j}}$, and
this can be computed readily by recurrence.


\item      {\sl Choice of impact parameter}

The second issue is by no means as simple.
In a single encounter between two stars of velocity $\bv_1, \bv_2$,
respectively,   square of the velocity change of the first star is  given by
$\displaystyle{(\Delta v_1)^2 = \frac{4G^2m_2^2}{p^2\vert\bv_1-\bv_2\vert^2}}$.
We compare this with the  mean square change in time $dt$ given by the
theory of relaxation, which is 
$\displaystyle{\langle(\Delta v_1)^2\rangle = \frac{8\pi
    G^2m_2^2n_2\ln(\gamma N) dt}{\vert\bv_1-\bv_2\vert}}$, where
$\gamma$ is a constant of order unity.  Therefore,
in order to mimic relaxation by a single encounter, in a statistical
sense, we  choose the impact parameter by the formula $p =
\displaystyle{\frac{1}{\sqrt{2\pi n_2\ln(\gamma N) dt\vert\bv_1 -
      \bv_2\vert}}}$.  This requires an estimate of the density $n_2$,
for which \henon\  devised the following procedure.
 The mean radial separation $dr$ between shells is given by $4\pi n_2
  r^2 dr = 1$, and this could be used as the basis of an estimate of
  $n_2$.  To reduce fluctuations, however, \henon\  chose the distance to the 5th nearest
  neighbour  to estimate $n_2$.

A couple of comments are due.  First, note that choosing encounters
between successive pairs of stars automatically ensures that  each
star encounters stars of mass $m$ in proportion to the local stellar
mass function.  Second, and more remarkable, is that this procedure
ensures not only the correct value of $\langle(\Delta v_1)^2\rangle$,
but also the correct value of all other first and second moments of
$\Delta v_1$. 

\item      {\sl Choice of new position}
  Between encounters, stars move on planar orbits in the continuous
  spherical potential $\phi(r)$ with energy $E$ and angular momentum $J$.
Thus
$ E = \displaystyle{\frac{1}{2}\left(v_r^2 + \frac{J^2}{r^2}\right) +
  \phi(r)}$, where $v_r$ is the radial  component of velocity.
The probability of finding the star between radii $r$ and $r + dr$ is proportional to the time spent there, i.e. $f(r)dr \propto\displaystyle{\frac{dr}{v_r}}$.
To choose $r$ with this probability density, one could in principle
use the rejection/acceptance technique.  However, this is applicable
only when the probability density $f$ is bounded, and here
$f(r)\to\infty$ as $v_r\to0$, i.e. 
at the extreme radii (pericentre and apocentre, denoted by
$r_{min},r_{max}$) on the orbit (Fig.\ref{fig:fofr}).
          \begin{figure} 
            \caption{(Left) A schematic of an orbit in a spherical potential.
              The radial component of velocity, $v_r$, vanishes at
 pericentre and apocentre, where  {the probability density of radius
   $r$ is infinite (Right).}
}           \label{fig:fofr}
  \begin{minipage}
    {4.cm}
	\begin{center}
	  \includegraphics[width=4cm,height=4cm]{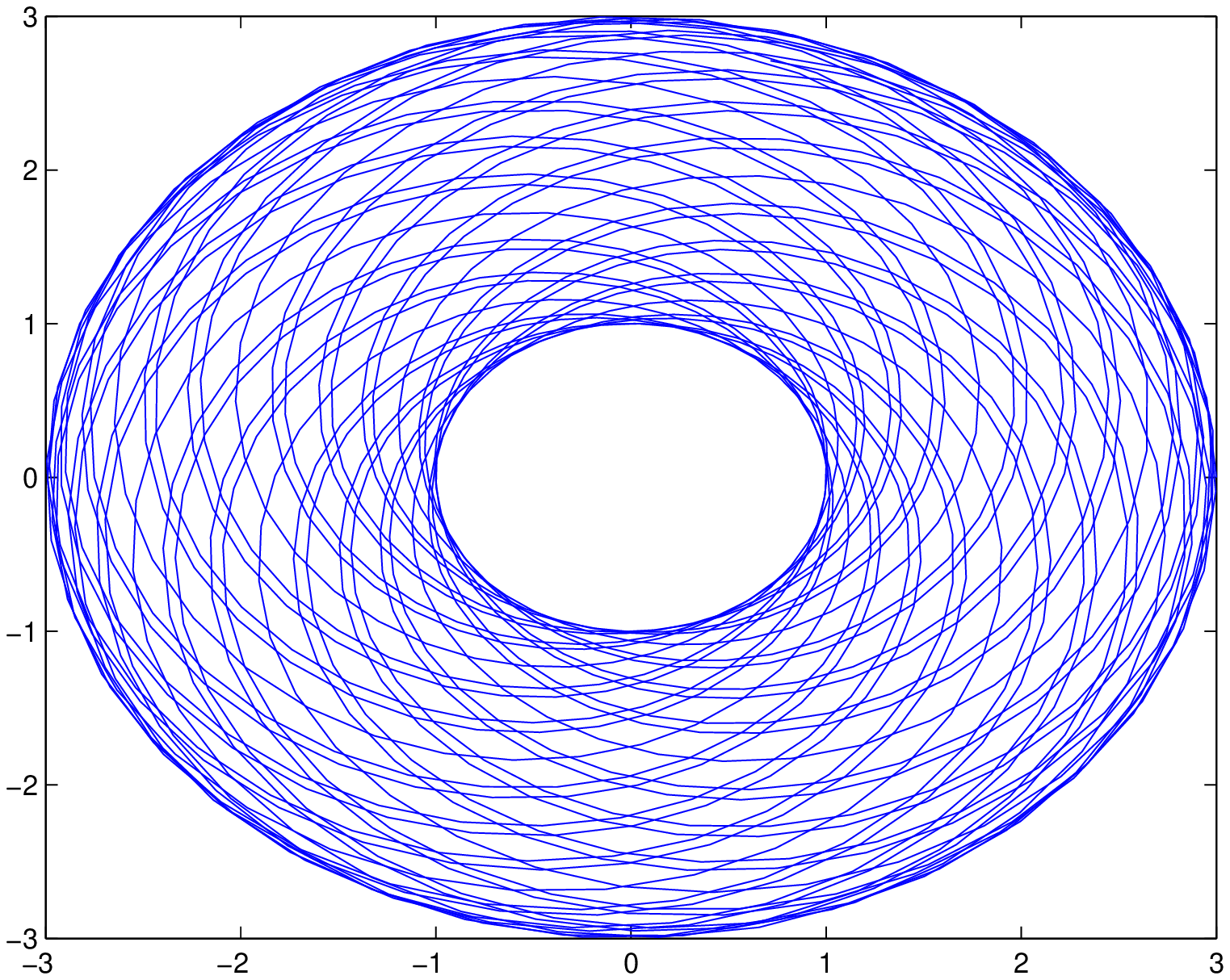}            
	\end{center}
  \end{minipage}
  \begin{minipage}
    {4.cm}
	\begin{center}
          \includegraphics[width=5cm]{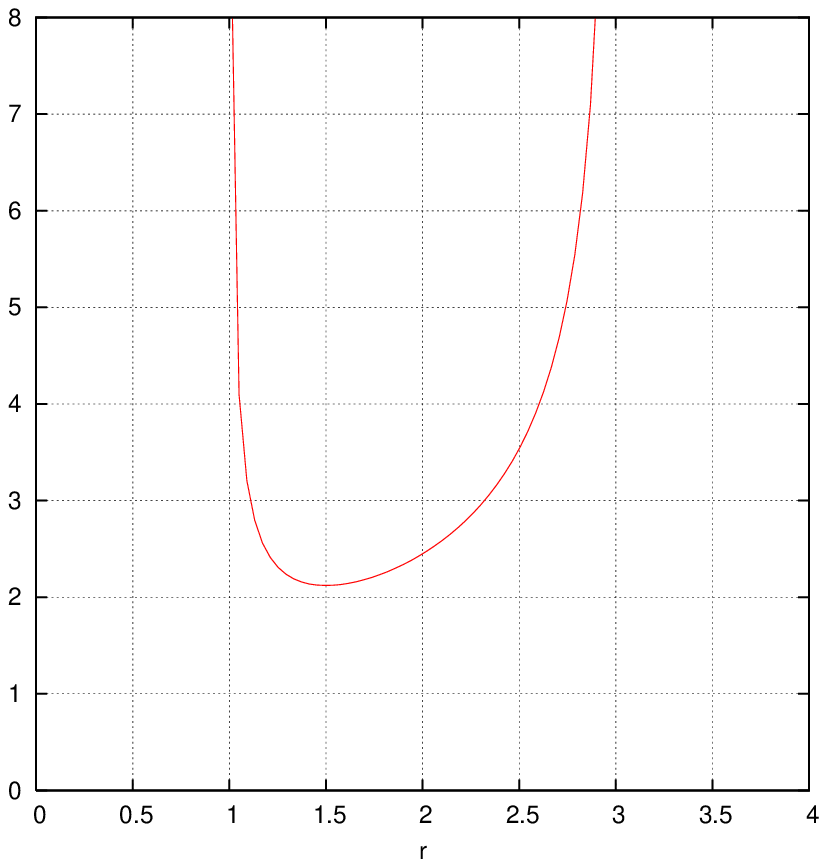}            
	\end{center}
  \end{minipage}
	  \end{figure}

\henons\  solution to this hurdle has not been superseded.  Following
him, we
 choose a new radial variable $s$, whose form is to be determined.
 Then $f(s) = f(r) \displaystyle{\frac{dr}{ds}}$.   Now we choose $s$ so that $\displaystyle{\frac{dr}{ds}}$ also vanishes
  (like $v_r$) at $r_{min}$ and $r_{max}$.  
In fact we      seek $r(s)$ so that $\displaystyle{\frac{dr}{ds}} = 0$
at $s=s_{min},s_{max}$  and such that $r_{min,max} = r(s_{min,max})$.
\henons\  choice was to set  $s_{min,max} = \pm1$ and  let
$\displaystyle{\frac{dr}{ds}} = 3A(s+1)(1-s)$, where $A$ is constant.
Hence $r(s) = B + A(3s-s^3)$, where $B$ is another constant.  Finally, choose $A,B$ so that $r(\pm1) = r_{min,max}$.

\subsection{Comments on the method}
\end{enumerate}


The time step $dt$ is of order the relaxation time, and   each step
takes of order $N\ln N$ operations (because sorting of the radii is
required, and computation of the potential at the new position of a
star requires $O(\ln N)$ operations).  Thus the   computational effort
is of order $N\ln N$ per relaxation time, by contrast with, for
example, a direct $N$-body simulation, where for a relatively
homogeneous system the effort is of order
$\displaystyle{\frac{N^{10/3}}{\ln\gamma N}}$ per relaxation time
\citep{1988ApJS...68..833M}.  

\begin{figure}
\caption{A scatter-plot of the Galactic globular clusters:  half-light
  relaxation time against absolute V-magnitude.  The data are taken
  from the most recent version of the Harris catalogue
  \citep{1996AJ....112.1487H}.  The absolute magnitude is taken as a
  proxy for $N$, assuming constant mass-to-light ratio and mean stellar
  mass.  The sloping lines represent systems requiring roughly the same
  computational effort with a modern Monte Carlo code, assuming the
  scaling given in the text.  Adjacent lines correspond to effort
  changing by a factor of 10, and lines for 1 and 10 days are
  labelled.
  One caveat is that both $t_{rh}$ and
  $M_V$ vary throughout the lifetime of the cluster, and it is assumed
that the current values may be adopted.}
\label{fig:trhMMC}
  \includegraphics[width=.5\textwidth]{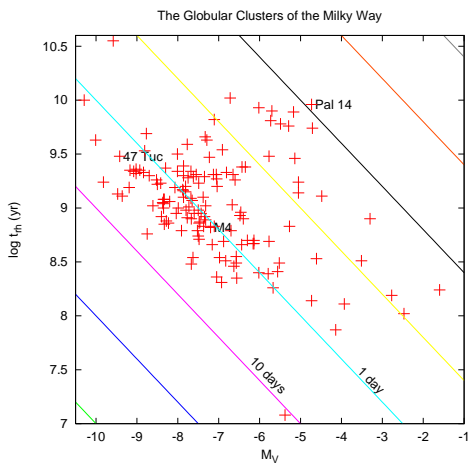}
\end{figure}

Clearly the Monte Carlo method wins for the range of $N$ corresponding
to the globular star clusters.  In fact
any Galactic globular cluster can be modelled in roughly 1 day (Fig.\ref{fig:trhMMC}), whereas
with direct $N$-body methods the modelling of a given cluster for a
Hubble time is
limited at present to $N\ltorder2\times 10^5$ stars\footnote{It is true that
  substantially larger simulations have been published, but these are
  either relatively short, or restricted to a single simulation,
  whereas finding the initial conditions appropriate to a given
  cluster requires many simulations, in order to ascertain optimal
  initial conditions.}.


Some of the merits and limitations of the Monte Carlo method are readily summarised.  Besides its
relative efficiency, just mentioned, we have the fact that one can
``readily'' add stellar evolution, and dynamical interactions
involving binaries.  Each required considerable effort in coding, but
the job is done.  Of course, these are not advantages relative to the
$N$-body method, but they do account for the fact that the Monte Carlo
method leads the field among the fast methods for the simulation of
dynamical evolution of rich star clusters.

Among the difficulties of the Monte Carlo method is the modelling of
the Galactic tide.  This has been steadily improved, but still
$N$-body methods are needed for an adequate modelling of such features
as the
kinematics of escaping stars \citep[e.g.][]{2010MNRAS.407.2241K}.
It also has to be assumed that the system is  spherically symmetric
and non-rotating.   Triples and higher-order multiples have to be
ignored, though further development of the code could quite readily
remove this limitation.

Because of the approximations involved, the Monte Carlo code has been
repeatedly compared with the results of the virtually assumption-free
$N$-body method, at least in the range of $N$ which the latter can
reach.  Naturally the first of these studies was carried out by \henon\  himself \citep{1974A&A....37..183A}. 

At present the Monte Carlo method is the method of choice for all 
the large globular clusters of the Galactic system, and has been
applied to the modelling of the individual objects
 $\omega$ Cen \citep{2003MNRAS.339..486G}\footnote{This study used
  scaled models, i.e. models in which the number of stars is much
  smaller than that in the actual cluster, though the radius is also
  changed to ensure that two-body relaxation proceeds at the correct rate.}, M4
\citep{2008MNRAS.389.1858H}, NGC6397\citep{2009MNRAS.395.1173G}, 47
Tuc \citep{2011MNRAS.410.2698G} and M22 \citep{2014MNRAS.439.2459H}.
It is also being used to study such topics as the origin and
distribution of blue stragglers \citep{2013MNRAS.429.1221H,2013ApJ...777..105S,2013ApJ...777..106C}.


 \subsection     {\sl Digression: $N$-body Units?}

About 30 years ago this author tried to encourage the community to
adopt a common system of units for its work, so that different studies
could be more readily compared.  It is 
defined by setting
\begin{eqnarray*}
  G &=& 1\\
M &=& 1\\
R &=& 1,
\end{eqnarray*}
where $G$ is the constant of gravitation, $M$ is the total mass of the
system, and $R$ is its virial radius. 
Though some researchers find these units inconvenient for some reason or
another, nevertheless they have been taken up widely, partly because
major $N$-body and Monte Carlo codes adopt them, and perhaps  also because they preserve
the sense that simple $N$-body models (i.e. without the accretion of
stellar evolution and so on) are scale-free.  As a result the paper in
which the author's recommendation appeared \citep{1986LNP...267..233H}
is ranked 5th by number of citations among this author's papers, even though the publication in
which it appeared was almost unobtainable for a long time.  Furthermore the
units were not even his idea, and so the paper's prominence has long
been a source of mild embarrassment.

 Actually, $N$-body units originated in H\'enon's papers on the
  Monte Carlo method (see especially
  \citealt{1971Ap&SS..14..151H,1972ASSL...31..406H}).  This fact was
implicitly  acknowledged in the author's paper, which did not even use
   the term {\sl $N$-body units}.  It seems to the author altogether appropriate
to recommend that they be referred to in future as  {\sl\henon\ 
  units}\footnote{This was actually done by one speaker at the
  Gravasco program which ran at the Institut Henri Poincar\'e from 9
  September until 13 December, 2013, a point which prompted the author's remarks.}.



\subsection{\henons\  Principle}\label{sec:principle}

This is another opportunity to commemorate \henon, but the idea which
he introduced is more profound than in the case of \henon\  units,
important though those units are in practice.

Simple collisional stellar systems go through a process of core
collapse, and for some time this was an impasse in the study of
such systems.  In the late 70s the fate of a star cluster {\sl after}
core collapse seemed utterly conjectural.  (See, for example, the
introduction to \citet{1978ApJ...219..629L}.)

  In 1974, post-core-collapse evolution could not be simulated
  with a Monte Carlo code.  \henon\ knew, from his thesis work
  (Sec.\ref{sec:thesis}), that binaries were the missing dynamical
  ingredient, but their introduction into the Monte Carlo code was
  only accomplished later \citep{1986AcA....36...19S}.  Their role is
  to act as a ``heating'' source, giving kinetic energy to the stars
  with which they interact.
\henon\ 
  realised, however, that the details of the heating were irrelevant,
  except for determining the parameters of the core.  He argued that the core adjusts to produce the
  energy required for ``balanced evolution'' of the system as a whole:
 if the core is too small and dense, it produces too much
    energy, and expands, which reduces the energy production, while
 if the core is too large and dilute, it produces too little
    energy, and contracts, which increases the energy production.
    \henon\  also knew what the ``required'' energy was.  It is 
the flux of energy from the
  centre in H\'enon's homological models (see Sec.\ref{sec:thesis}).
This idea, that the core responds to the energy requirements of the
cluster as a whole, can be referred to as ``H\'enon's Principle''.  It
is analogous to Eddington's realisation \citep[][especially ch
  1.4]{1926ics..book.....E} that one can predict the luminosity of a star even without knowing the
nature of the source of stellar energy. 

\henons\  Principle has several important applications, and is one of
the foundation stones of our understanding and simulation of the
long-term evolution of star clusters.   To allow a Monte Carlo model
to pass through core collapse, \henon\ took the bold step
\citep{1975IAUS...69..133H} of introducing a
quite artificial energy-generating mechanism, affecting only the
innermost particle in the model, but in a manner mimicking the
behaviour of a real core.  In this way  core parameters will be wrong,
but the overall evolution will be correct.

The value of \henons\  Principle extends well beyond the Monte Carlo
model, however.  It applies if the density-dependence of the
  energy-generating mechanism is high enough that almost all energy is
  generated in the core.  For example,  three-body binary formation gives a rate of energy
    generation proportional to $\displaystyle{
\frac{G^5\rho^2m^3}{\sigma^7}}$ per unit mass, where $\rho,\sigma$ are
    the density and one-dimensional velocity dispersion, respectively;
    from which \henons\ 
    Principle allows one to estimate the radius and density  in the
    core of the cluster \citep[see, for example,][]{1987ApJ...313..576G}.
In a similar way, core parameters may be estimated  in terms of the overall structure of the cluster (total
  and mean particle mass, half-mass radius) if the energy-generating
  mechanism is dynamical evolution of primordial binaries \citep{1994ApJ...431..231V}, stellar interactions with a 
central black hole \citep{2007PASJ...59L..11H}, and perhaps even  stellar evolution \citep{2013ASPC..470..339G}.




 


\section{The homological models (1961-1965)}\label{sec:thesis}

\subsection{The 1961 paper}\label{sec:henon1961}
{}
We go back about 10 years to two papers which laid the groundwork for
\henons\  Principle, but also did much more \citep{1965AnAp...28...62H,
1961AnAp...24..369H}.  His principal aim was to solve the isotropised
Fokker-Planck equation for the evolution of a star cluster, i.e. the equation

\begin{equation}
      \begin{aligned}
         Q'\frac{\partial F}{\partial T} -    F'\frac{\partial Q}{\partial T}
=
\frac{\partial}{\partial
      E}\left[F\int_{-\infty}^EF_1Q_1'dE_1\right. +&\\
+\left.F'\left(\int_{-\infty}^EF_1Q_1dE_1 + Q\int_E^\infty F_1dE_1\right)\right]&,
      \end{aligned}
\label{eqn:fpe}
  \end{equation}
where
 $F(E,T)$ is the 1-particle distribution function, expressed as a function of
  particle energy $E$ and time $T$,
 $Q(E,T)$ is the phase-space volume up to energy $E$ at time $T$,  $'$
  denotes the energy-derivative
  $\displaystyle{\frac{\partial}{\partial E}}$, and  the variable of
  integration appears in the forms $F_1 = F(E_1,T)$ and $ Q_1 = Q(E_1,T)$.


Before tackling this formidable problem head-on, \henon\  examined it
from several angles.  First, assuming that the central density is finite, he computed one
      time step, and showed that the central density always increased.  He
      inferred that solutions might have infinite central density.
      Next, his review of the literature pointed out that previous
      work had
  either (i)
 neglected the collision term (the right-hand side), and hence
    approximated the Fokker-Planck equation with the Collisionless
    Boltzmann equation; or (ii)
 assumed a steady spatial structure (e.g. a square well potential) and
  solved the collisional evolution, thus neglecting a term on the left
  hand side.  \henon\   asserted that {\sl all} terms played a
  comparable role, and  decided to search for self-similar (or
  homological) solutions.  Much later, Lynden-Bell explained why it is
  that self-similar solutions are of such importance and relevance in
  such problems \citep{1980MNRAS.191..483L,1983MNRAS.205..913I}.  The
  point is quite subtle, but it least it is easy to see that the
  solving for such a solution simplifies the task a lot.


{}
 
Even so, \henon\  needed to bring his powerful technique and intuition
to bear, especially with regard to the question of boundary conditions.
Deep in the cluster, he noted that the collision term, i.e. the
right-hand side of eq.(\ref{eqn:fpe}) must nearly vanish.  Exploiting
some simple freedom of scaling, this implies that, to lowest order,
the distribution function is a Boltzmann distribution, i.e.
  $F = \displaystyle{e^{-E}}$.  Next he obtained an improved
approximation (still with the assumption that the collision term
vanishes) by  posing $F = \displaystyle{e^{-E}} + \delta F$.  In this
way he
showed that  
\begin{equation}F = \displaystyle{e^{-E}} + Ke^{-E/2} +
  K_2(5-E)e^{-E/2},\label{eqn:ibc}\end{equation}
 and by examining the mass and energy of part of
  the system, he showed how the new terms could be interpreted:  
the term in $K$ corresponds to an energy flux at the centre ($R=0$),
 while the  term in $K_2$ corresponds to a mass flux at $R=0$.  \henon\ 
 knew that  $N$-body simulations \citep{1960ZA.....50..184V} had
 already revealed the
  formation of energetic binaries.  On such grounds \henon\  allowed
  $K$ to be non-zero in general, while he assumed that  $K_2 = 0$.

  Thus conditions near the centre of the system are fixed.
 The earlier of the two papers we are discussing
 \citep{1961AnAp...24..369H} dealt with a star cluster immersed in the
 gravitational field of a parent galaxy, which fixed the boundary
 conditions at the outside.  This left \henon\  with the task of solving
an integrated form of the isotropised, self-similar Fokker-Planck equation, i.e.
  \begin{eqnarray*}
\frac{3}{2}b\int_{-\infty}^EF_1Q_1'dE_1 &-& 3bFQ 
=
F\int_{-\infty}^EF_1Q_1'dE_1+\\
&+&F'\left(\int_{-\infty}^EF_1Q_1dE_1 + Q\int_E^\infty F_1dE_1\right),
  \end{eqnarray*}
where $b$ is another constant to be determined, which arises from the
time-dependent terms in eq.(\ref{eqn:fpe}).  This equation is
equivalent to a fourth-order system.  The inner boundary
conditions are given by the expressions 
$F\sim\displaystyle{e^{-E} +
  Ke^{-E/2}}, E\to-\infty$, and three similar expressions for the
other three variables; and the outer boundary conditions are
 $F = 0$ and  $\displaystyle{\int_E^\infty F_1dE_1
  = 0}$ at $E = 0$, since $F=0$ for $E>0$, if the potential vanishes
at the tidal boundary.  (Note that two of the four variables in the
equivalent system are undetermined at the outside.)  Auxiliary equations to be evaluated or solved are
for the  density $D = \int_U^0(2E - 2U)^{1/2}F dE$; for the potential
$U$, i.e. Poisson's equation $\displaystyle{\frac{d^2Z}{dU^2} = -
  D\left(\frac{dZ}{dU}\right)^3 Z^{-4}}$ where $Z = 1/R$; and the
phase-space volume $Q =
\frac{1}{3}\int_{-\infty}^E(2E-2U)^{1/2}R^3dU$.  Given a fourth-order
system with four unknowns (the constants $K$ and $b$, and two outer
values), it is reasonable to suppose that an isolated solution might
exist, and no doubt this is one reason why \henon\ chose to retain one
of the terms in eq.(\ref{eqn:ibc}).


{H\'enon's solution method was iterative:}
  \begin{enumerate}
  \item Guess $F,K,b$
\item Do
  \begin{enumerate}
  \item Compute $\rho,U,Q$
\item Solve the Fokker-Planck equation for $F$, and ``t\^atonner''\footnote{``To feel one's
    way, to grope; to proceed tentatively'' \citep{Ba1927}.} on 
  $K,b$ to satisfy the outer boundary conditions.  This gives new $F,K,b$.
  \end{enumerate}
\item Until ($F$ converges)
  \end{enumerate}

Remarkably enough the method did indeed converge.  Indeed the changes in the
unknowns decreased by a factor of about 5 in each cycle of the main
loop.  The calculations were performed on an
  IBM 650 at Meudon, and in  8 hours the  solution was obtained
 with an estimated precision of 1 part in 1000.


\henon\  presented his results in the form of both graphs and tables, giving
$F$ and the other useful functions obtained by his computations.
Table \ref{tab:henon1} gives the two essential unknown parameters, $K$
and $b$, along with results from a recent study, still ongoing.
    \begin{table}
      \caption{Parameters of the  homological models}
  \begin{tabular}{lcc}
\label{tab:henon1}
    &K&b\\
(i) Tidally limited model\\
\citet{1961AnAp...24..369H} &-0.9400&-0.2719\\
Apple, Heggie, Mackie \& \\
~~~Walters (in preparation)&-0.9176&-0.2763\\
(ii) Isolated model\\
\citet{1965AnAp...28...62H}&-1.363&-0.7313\\
Apple, Heggie, Mackie \& \\
~~~Walters (in preparation)&-1.330&-0.7583
  \end{tabular}
    \end{table}
The latter study used matlab, and  also takes a few hours, but it includes
a  lengthy investigation of the dependence of the results on the
tolerances of the numerical methods, and on the choice of the lower
end of the range of energy $E$, as the domain of the mathematical
problem is semi-infinite.  Examining the convergence of our results as
these choices are varied, we conclude that the results are accurate to
the last significant figure in the table.

We have only just touched on the numerous topics which \henon\  treats
in this paper.  One of these is the evolution of two-component star
clusters, in which he made the approximation of assuming that one
component contributes negligibly to the total density.  An attempt to
remove this approximation was the main motivation for the recent work
by Apple et al, discussed above.  Actually, it seems unlikely that
such a model exists, as all the evidence points to the more rapid
escape of low-mass stars across a tidal boundary, whereas in a fully
homological model, the ratio of the total mass in both components must
be constant.  For an isolated Fokker-Planck model, however, the mass
is constant (Sec.\ref{sec:escape}), and so the search for a
homological, two-component model seems more promising.

\begin{figure}
\caption{The
      reception of H\'enon's 1961 paper (ADS), measured by  the number of refereed
and unrefereed citations each year.  The data are not complete.  For
example, the citation in \citet{1965AnAp...28...62H} is not in ADS.
}
\includegraphics[width=.5\textwidth,clip=true,trim=0 0 0
  0]{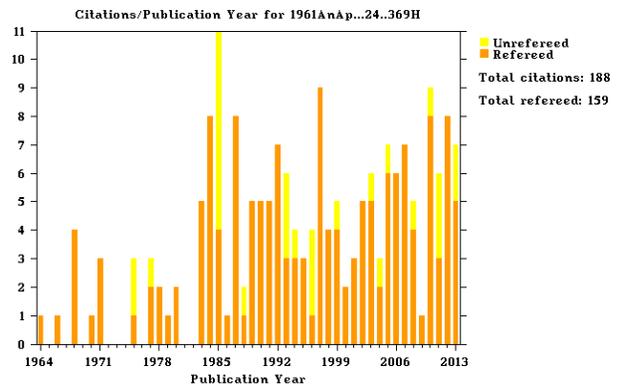}
\label{fig:henoncites}
\end{figure}

Though the 1961 paper contained such a wealth of new material --- and
there is much more than we have reviewed here --- its
reception may have been a disappointment to \henon.  It was scarcely
cited in the decade following its publication (see
Fig.\ref{fig:henoncites}), but thereafter it has taken its place as
part of the foundations on which our understanding of star
cluster evolution is built.




\subsection{The 1965 paper}


  


There can hardly be any greater contrast than that between the 1961 and
1965 papers.  The earlier paper is very long, but the later paper is
terse, to the extent that not even the numerical method is discussed,
only the results.
At first sight, the later paper deals with a less realistic problem
than the 1961 paper, as the star cluster is regarded as {\sl
  isolated}.  However, we shall see that it is needed for a complete
understanding of star cluster evolution, even for star clusters
immersed in a parent galaxy.

Among the technical differences between the two cases are the details
of the  Fokker-Planck equation to be solved, because the
time-dependence of the scaling of radius and mass is different.  The
inner boundary conditions are as before, however, including the flux
of energy at the centre.  The outer boundary conditions are not stated
by \henon, and one difficulty experienced by Apple et al in recently
revisiting these problems was to find out what they are and how to apply them.

Following \henon, we assume that the potential vanishes at infinity
and that $F(E)= 0$ for $E>0$.
Then the integral $\int_E^\infty F(E_1)dE_1$, which appears in the
Fokker-Planck equation, tends to $0$ as $E\to0-$.  Now it
can be shown from the Fokker-Planck equation that, necessarily,
$F(E)\to 0$ as $E\to0$.  In the tidally bound problem, however, these
are independent conditions, and so in the isolated case the latter has to be replaced by
another condition, for which Apple et al take the stated condition
$U\to0$ as radius $R\to\infty$.  

This  change of outer boundary data requires a different solution
strategy, which can be summarised as follows:


  \begin{enumerate}
  \item guess $F,K,b$
\item do
  \begin{enumerate}
  \item compute $\rho,U,Q$
\item solve the Fokker-Planck equation for $F$, and iterate on
  $K$ to satisfy $\int_E^\infty F(E_1)dE_1\to0$ as $E\to0-$
  \item iterate on $b$
  \end{enumerate}
\item until $U\to0$ as $R\to\infty$
  \end{enumerate}

With modern software all ``t\^atonnement'' could be avoided.  The
results are compared with \henons\ in Table \ref{tab:henon1}.

  \subsection{Application to the Globular Clusters of the Milky Way}

To illustrate the influence which \henons\  papers have enjoyed in
recent years, we summarise here a model for the evolution of the
globular clusters in our own Galaxy.  First of all let us compare the
time-dependence of the evolution of the two models.
    

\begin{table}
  \caption{Comparison of the isolated (1965) and tidally truncated
    (1961) models}
\label{tab:time-dependence}
    \begin{tabular}{l|l|l}
\hline
    &Isolated &Tidally truncated \\
\hline
Radius&  $\displaystyle{ R(0)\left(1 +
  \frac{T}{T_r(0)}\right)^{2/3}}$&  $\displaystyle{ R(0)\left(1 -
  \frac{T}{T_r(0)}\right)^{1/3}}$\\
& expands & contracts\\
\hline
Mass&  $M(0)$ &  $\displaystyle{M(0)\left(1 - \frac{T}{T_r(0)}\right)}$\\
& No mass loss & Mass decreases\\
\hline
Relaxation  &  $\displaystyle{T_r(0)\left(1 + \frac{T}{T_r(0)}\right)}$&
$\displaystyle{T_r(0)\left(1 - \frac{T}{T_r(0)}\right)}$\\
~~~time&Increases&decreases\\
\hline
  \end{tabular}
Note: some numerical constants in the relaxation time have been omitted.
\end{table}



We suppose that the initial radius of a cluster is much less
than its tidal radius.  Then the cluster expands (much like \henons\ 
isolated model) until its radius
  becomes comparable with the tidal radius.  After that it evolves much
  like the
  tidally limited model.  In fact it is possible to devise a unified
  model which smoothly goes over from one form of evolution to the
  other \citep{2011MNRAS.413.2509G}.  The only  modification we make
  is to assume that the evolution is faster than in \henons\  one
  component models, because a
  range of stellar masses is present in more realistic clusters.


To illustrate the consequences of this model, we note that, in the
first (essentially isolated) phase, the relaxation time behaves as
$T_r(T) = T_r(0) + T$ (Table \ref{tab:time-dependence}, noting the
footnote there).  If the  cluster was initially  very compact,
$T_r(0)\ll T$, and so $T_r(T) \simeq T$.  Finally, since  all clusters
have nearly the same age, it follows that all have the same relaxation
time, of order their age.

\begin{figure}
    \caption{The survival triangle of Galactic globular
  clusters, from \citet{1997ApJ...474..223G}.  It is a scatter plot of
  the mass and half-mass radius of the clusters, differentiated by
  their current galactocentric distance $R$.  At lower right the
  destruction time scale due to disk- and bulge-shocking becomes
  comparable with the clusters' age; at the top clusters would be
  destroyed by dynamical friction in the Galaxy; and at lower left
  destruction is dominated by two-body relaxation: the slope of the
  linear part of the outer contour corresponds to a line of constant
  relaxation time.  The smooth corners
  of the triangle come from adding the destruction rates of the three
  processes, and different contours are obtained at different $R$.}
\includegraphics[width=.5\textwidth]{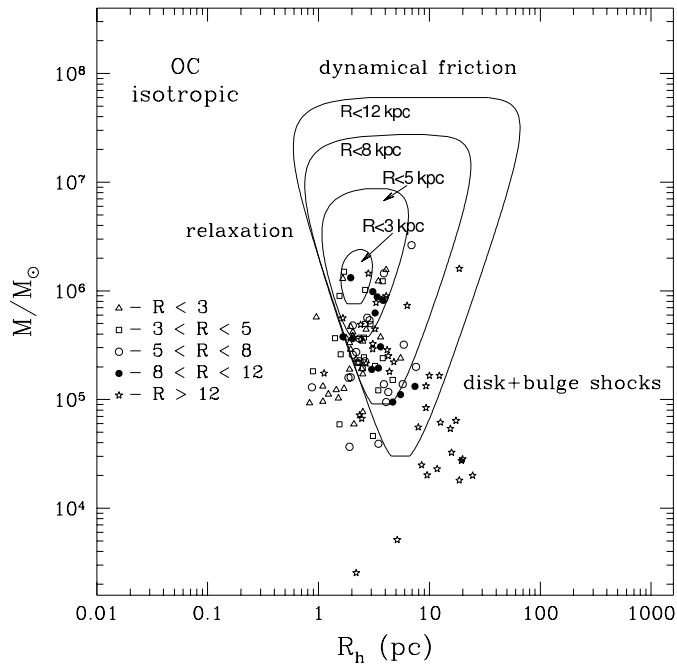}
\label{fig:survival}
\end{figure}

Let us compare this result with the well known ``survival triangle'',
shown in Fig.\ref{fig:survival},
which seeks to establish that the distribution of cluster parameters
is restricted by the time scales on which various destruction
mechanisms act \citep{1977MNRAS.181P..37F,1997ApJ...474..223G}.  The
usual interpretation of this diagram is that it shows that clusters
are not found where the mechanisms of destruction would remove them
within their lifetime.    While most clusters sit comfortably inside
the outermost contour in this diagram,
  they concentrate to the bottom left, where their distribution
  appears to be limited by destruction due to two-body relaxation.
  But  one can perhaps better
  interpret the diagram as showing that most clusters have almost the
  same relaxation time, as predicted in the model of \citet{2011MNRAS.413.2509G}.

\section{Escape (1958-1969)}\label{sec:escape}

\subsection{Isolated star clusters}\label{sec:isolated}

This section covers a topic which, unlike the foregoing, probably does
not have any evident application to the star clusters.  It is a
question of ``pure'' stellar dynamics, but a loose end which \henons\ 
work did a lot to expose.  It begins with his work on the diffusion
coefficients in the theory of two-body relaxation (see the
contribution in this volume by F. Namouni).  A side-product of this work
was an expression  for 
the probability that, in time interval $dt$, a star experiences an
encounter which changes its  velocity  from $\bV$ to $\bV + \be$, namely
$$
P = \frac{8\pi G^2 m^2 dt}{e^5}d^3\be\int_{v_0}^\infty a(v) vdv,
$$
\citep{1960AnAp...23..467H}, where
 $v_0 = \displaystyle{\frac{1}{e}\left|\bV.\be + \frac{M+m}{2m}e^2\right|}$,
 $m,M$ are the stellar masses, and 
 $a(v)$ is the distribution of velocities.  \henon\  quite rightly
referred to  ``le r\'esultat final, d'une remarquable simplicit\'e''.
Indeed it is, by comparison with the intricacies of its derivation,
but this author knows no simpler derivation than \henons, and indeed
the result seems unattainable if one does not follow his footsteps
closely.  

In the above paper, he used this result to obtain expressions for third-order
moments of the velocity change $\be$, verifying the known result that
these are smaller than the first and second moments (i.e. those that
appear in the Fokker-Planck equation) by a factor of order the Coulomb
logarithm $\ln\gamma N$, where $N$ is the total number of stars.



A second application of this result, obtained by integrating over speeds
$\vert\bV + \be\vert$ above the escape speed, and over all stars, is
an expression for the rate of escape from an isolated stellar system, i.e.
$$
\frac{dN}{dt} =
-\frac{256}{3}\sqrt{2}\pi^4G^2m^2\times~~~~~~~~~~~~~~~~~~~~~~~$$
$$
\times\int_0^{\infty}r^2dr\iint_{E+E'-U>0}\frac{f(E)f(E')(E+E'
  - U)^{3/2}}{E^2}dE dE'
$$
\citep{1960AnAp...23..668H}
where
 we assume the potential $U(r)\to0$ as $r\to\infty$,
and $f(E)$ is the distribution function expressed as a function of
  energy $E$.
As a specific example, \henon\  calculated the result for a  Plummer
model, obtaining the result $\displaystyle{\frac{dN}{dt} =-
  0.00426\sqrt{\frac{GmN}{r_0^3}}}$
where
 $r_0$ is the projected half-mass radius.  This result has been
checked by $N$-body simulations \citep{1994MNRAS.268..257G}.  \henon\ later extended this
result to the case of a Plummer model in which stars may have
different masses \citep{1969A&A.....2..151H}.

An interesting point to note here is the absence of any  Coulomb
logarithm.  By contrast, most existing theories for the escape rate in
an isolated system assumed that the part of the velocity distribution
above the escape speed would be refilled on a relaxation time scale
\citep{A1938}.  In that case the Coulomb logarithm $\ln\gamma N$
appears in the numerator of the corresponding formula for the escape
rate.  A further contrast is with \henons\  isolated self-similar
solution \citep{1965AnAp...28...62H}, where the escape rate is zero.
\henon\  explained these contradictions by showing, in an appendix to
the 1960 paper \citep{1960AnAp...23..668H}, that the escape rate
would vanish in the diffusion picture on which the theory of
relaxation is based.   As the star approaches the escape energy, he
argued, its
orbital period tends to infinity, and the time it spends in the denser
parts of the cluster, which is where relaxation is effective, becomes
an ever smaller fraction of the period.  Thus relaxation stalls,
explaining qualitatively why Ambartsumian's argument fails, and why
the rate of escape vanishes in the Fokker-Planck approximation.
Furthermore, the higher moments of energy changes, which are neglected
in this approximation, are smaller by a factor of the order of  $\ln\gamma N$, and so it is conceivable that these neglected terms could
give rise to a non-zero escape rate, on a time scale longer than the
relaxation time by a factor of order $\ln\gamma N$, as \henon\  found.


 There is one possibility for restoring the idea that escape takes
 place on the relaxation time scale, at least in the long term.  After
 core collapse, 
an isolated cluster in expansion is powered by binaries, as we have
discussed in Sec.\ref{sec:henon1961}.  The rate of energy generation is 
$\dot E\sim -\dot M\phi_c$, where $\phi_c$ is the central potential
\citep{1987ApJ...313..576G}.  Then 
  ``H\'enon's Principle'' (Sec.\ref{sec:principle}) shows that $\dot E\sim
  -\displaystyle{\frac{E}{T_r}}$, where now $E<0$ is the total energy
  of the cluster, and so $\dot M \sim -\displaystyle{\frac{E}{\phi_cT_r}\propto\frac{M}{T_r}}$.
 This is not inconsistent with \henons\  argument that the escape rate
 vanishes in the Fokker-Planck approximation, as the processes of
 binary formation and hardening are not included in this equation.    

\begin{figure}
  
  \caption{Fraction of the bound mass as a function of time for
    isolated $N$-body models.}
\includegraphics[width=.5\textwidth,clip=true,trim=0 0 0 0]{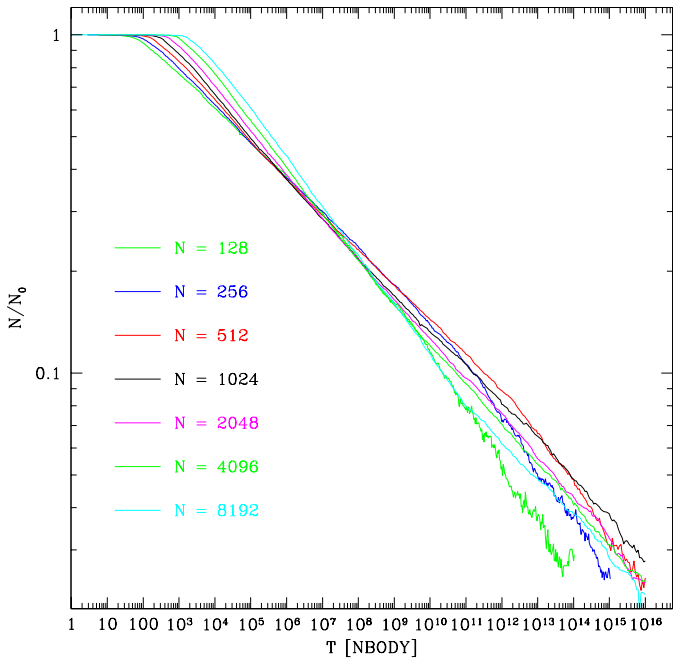}  \label{fig:baumgardt}
\end{figure}

One would have thought that numerical experiments would have settled
these questions, if indeed questions remain.  Unfortunately they
themselves raise fresh questions.  Simulations of enormous length are
needed, because the escape rate from an isolated system is so low, and
the expansion after core collapse slows down all dynamical processes.
The expansion also speeds up numerical simulations, however, and a
series of such simulations was carried out by H. Baumgardt
\citep{2002MNRAS.336.1069B} for $N$ up to a relatively modest 8192
particles, but the longest runs covered no less than about $10^{16}$
\henon\  units (Fig.\ref{fig:baumgardt}).  

Fig.\ref{fig:baumgardt} looks quite at odds with any of the foregoing
theoretical discussion, as it seems that all systems lose  about 75\%
of their mass in the same time, independent of $N$.  But two things
contribute here:  escape of stars, of course, but also the
post-collapse expansion, which controls all the time scales.  The
result in this figure could be understood qualitatively if the
large-$N$ systems expand more slowly (with reference to the relaxation
time) than small-$N$ systems.  Then they spend a larger time at radii
where the escape rate is larger.  Indeed Baumgardt et al find that
this is the case, and suggest that the reason for this is that the
clusters of different $N$ are not exactly scaled versions of each
other.  It is known, for example \citep{1987ApJ...313..576G}, that the
core radius is relatively smaller for larger $N$ in steady
post-collapse evolution.  Also, Baumgardt et al point out that there
are other escape mechanisms beyond the two-body process discussed by
\citet{1960AnAp...23..668H} and the three-body process discussed
above.  Lightly bound outlying  stars can
become unbound because of the recoil of the cluster due to escapers
resulting from the processes already mentioned.  In addition, these
slightly 
bound outlying stars can escape because
the potential well in which they sit becomes shallower, because of
escape by any of  the other processes already noted.  

This problem of escape from an isolated cluster, which initially seems so simple, raises a number of
tricky questions.  In principle some of them could be answered by
studying simulations, provided that there are not systematic
($N$-dependent) errors in the latter.  Certainly there is something to
be understood, because the result of Fig.\ref{fig:baumgardt} cannot
continue for arbitrarily large $N$; the duration of the first part of
the curve appears to be approximately proportional to $N$, and for
large enough $N$ the curve could not then drop steeply enough to cross the
approximate common intersection point of the curves shown.

  \subsection{Escape in a tidal field}

When a star cluster moves on a circular galactic orbit, it is immersed in the tidal field of the galaxy, and
 the basic issues of escape seem clearer.  \henons\  1961 homological
 model (Sec.\ref{sec:henon1961}) loses mass on the time scale of the
 relaxation time, though the tidal radius was treated as a cut-off,
 and the effect of the tidal field on members of the cluster, or on
 escapers, was ignored.

\begin{figure}
   \caption{Periodic orbits in Hill's problem, from
     \citet{1970A&A.....9...24H}. Jacobi constant $\Gamma$ increases to the
     left, and the ordinate is the coordinate, on the axis pointing to
     the centre of the galaxy, where an orbit cuts this axis.  The
     potential well is represented by the two bulges at upper right,
     with the Lagrange points L$_{1,2}$ at largest $\Gamma$.  The
     ``characteristic curve'' of family $f$ of periodic orbits extends
     from lower left, surrounded by the zone of quasi-periodic
     orbits.  They enter the physical extent of the cluster at about
     the middle of the diagram, but with an energy below the escape
     energy when the family passes the energy of the Lagrange points.}
\includegraphics[width=.5\textwidth]{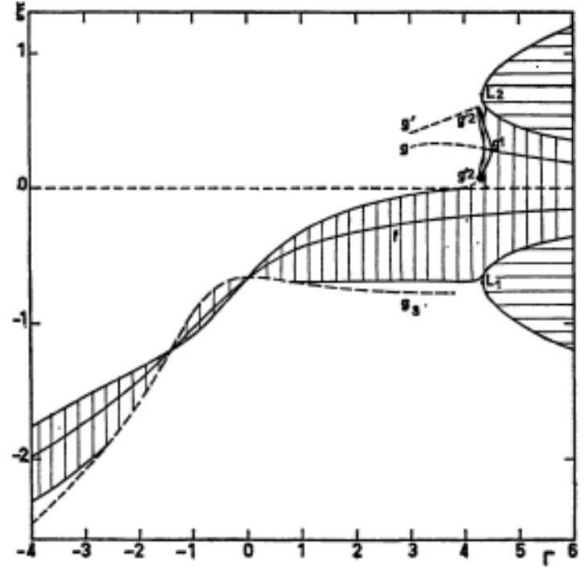}\label{fig:hillpo}
 \end{figure}

If the tidal field is properly modelled \citep{1942psd..book.....C},
the tidal field and inertial forces enter into the equations of
motion.  In fact the equations resemble those of Hill's problem in
celestial mechanics, a subject which \henon\  also studied as part of
his work on (mostly periodic) orbits \citep{1970A&A.....9...24H}.  He
was well aware of the possible implications for star clusters.  In
particular, there is a family, family $f$, of periodic orbits
(Fig.\ref{fig:hillpo})  which,
at large distance from the cluster, are essentially epicycles
governed by the tidal and inertial forces.   The cluster acceleration
is a perturbation, which actually acts to make the orbits stable.
\henon\  noted that a cluster could be surrounded by such orbits, though
he expressed some doubt whether they could be occupied by cluster
members in the normal course of cluster evolution.

 This family continues at lower ``energies''
(actually, lower Jacobi constant), and eventually is located {\sl
  inside the cluster}, even though the energy is still above the
 energy of escape.  In recent years such stars have become known as
 {\sl potential escapers}, though they could only every become
 escapers as a result of gravitational encounters, or because the
 potential well of the cluster changes by other processes.  These
 potential escapers can become a significant population, amounting to 
 10\% for $N = 16384$, and decreasing with $N$ only as $N^{-1/4}$ \citep{2001MNRAS.325.1323B}, so that they represent several percent even for real star
 clusters, provided that the approximation of a circular galactic orbit
 is adequate.  

The presence of potential escapers has some notable effects.  First,
models of star clusters never include this population, and it is not
known how this affects inferences made on the basis of such models
(e.g. estimates of the total mass).  It may well be that potential
escapers would look like stars with speeds above the escape speed, if
compared with a model in which such stars are ignored.  Certainly,
they have an important role in shaping the velocity dispersion profile
near the tidal radius \citep{2010MNRAS.407.2241K}.  
More indirectly, they
also act as a buffer between bound members (inside the cluster, and
with energies below the escape energy) and escapers, altering the time
scale of escape from about the relaxation time scale 
 $T_r$ to roughly $N^{-1/4}T_r$ \citep{2001MNRAS.325.1323B}, for
clusters filling the roche lobe (bounded by the Lagrange points).




Where the problem of escape still leaves the most tricky theoretical
questions is the case of  clusters on oval galactic orbits.
Apparently, such clusters lose stars at about the same rate as a
cluster on a circular orbit at some     intermediate radius between
the apo- and pericentres of its galactic orbit, at least when the
eccentricity of the orbit is not very large, and that the time
scale of escape varies with $N$ and $T_r$ in the same way \citep{2003MNRAS.340..227B}.
But a theoretical understanding of this problem is lacking.


\section{Epilogue}\label{sec:epilogue}

\begin{figure}
    \caption{Michel \henon\  in 1974, on an excursion made by participants
    at IAU Symposium 69 in Besan\c con.  Unfortunately he was in
    shade, and some distance from the camera.  Though the image has
    been heavily processed, for the author it is a vivid memento of
    Michel.}
\includegraphics[width=.5\textwidth]{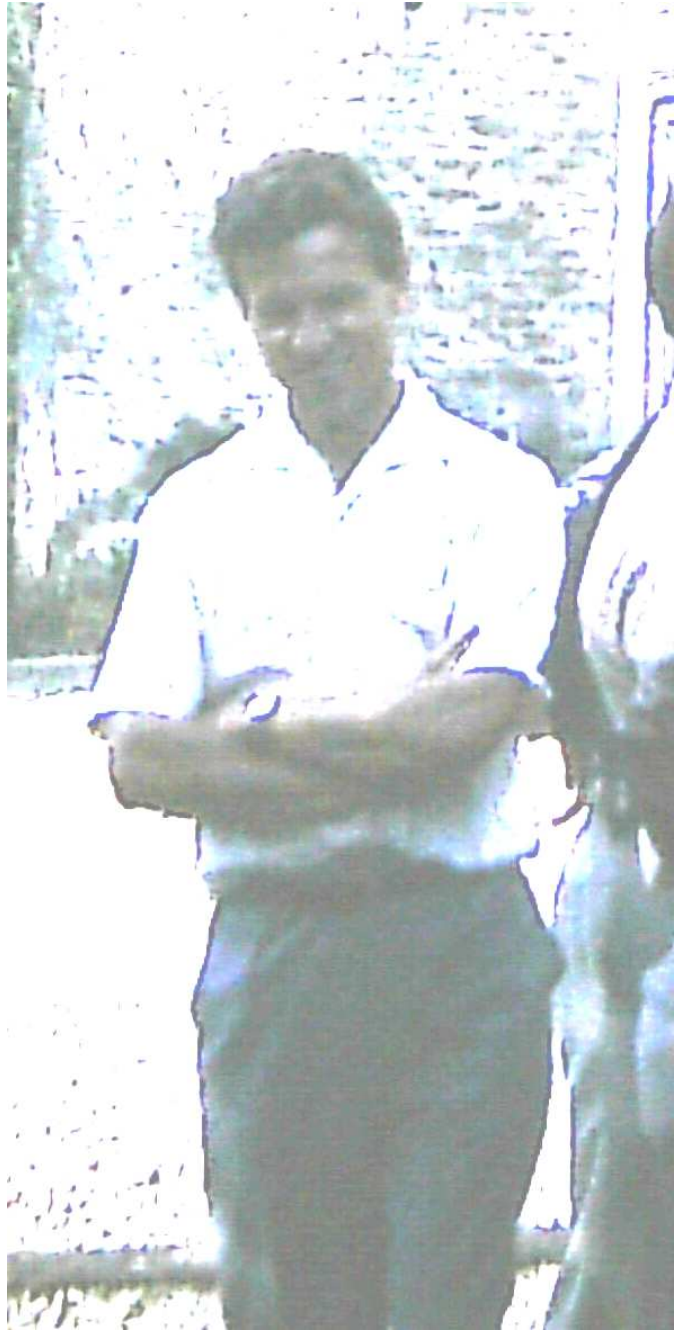}        \label{fig:henon}
\end{figure}

I met
\henon\  at several conferences (Fig.\ref{fig:henon}), but in 1975 I
also had the privilege of being hosted by him in my first foreign
postdoc.  For someone with little French this might have been a
challenge, and indeed during the spirited and noisy lunches at the Nice
Observatory it was.  But Michel's English was excellent, and
scientific exchange with him was straightforward.  In fact he told me
once that he preferred to use English for this purpose, 
as he found that some foreign
visitors who insisted in trying out their French simply made progress
next-to-impossible.

Though I thus returned to the UK knowing even less French than when I set
out for Nice, I picked some French up as the years passed, and a significant
amount I learned from \henons\  two papers on the homological models
(Sec.\ref{sec:thesis}).  These seem to me  a model of scientific
writing in any language.  Of course the language is broken by numerous
equations, but what makes the papers a pleasure to read is fundamentally
the quality of the scientific narrative.




The abiding importance of these papers led to a proposal for their translation into English.  This was done in autumn
2010 by Dr Florent Renaud, who had been working with Dr Mark Gieles on
an application of the models, and the translations are now in the public domain
\citep{2011arXiv1103.3499H,2011arXiv1103.3498H}.

Before setting about his work, Florent wrote to Michel, seeking his approval.
Here is his response.

\begin{quotation}
{\footnotesize      Cher coll\`egue,

Je vous remercie de votre proposition de traduction de deux de mes
articles, et bien entendu je vous donne mon accord enthousiaste.

L'article de 1961 constituait ma th\`ese de Doctorat. A l'\'epoque cette th\`ese
devait obligatoirement \^etre d'une pi\`ece et ne contenir que du texte
original, pas encore publi\'e (au lieu de consister en une collection
d'articles d\'ej\`a publi\'es, avec un peu de ``liant", comme cela se fait
maintenant). D'autre part ce texte devait \^etre enti\`erement en fran\c{c}ais.
Cela faisait partie du combat d'arri\`ere-garde men\'e par la langue fran\c{c}aise
contre l'anglaise! Je dois dire qu'apr\`es ma th\`ese j'ai continu\'e \`a publier
en fran\c{c}ais pendant encore quelque temps, avant de r\'ealiser que c'\'etait le
meilleur moyen de ne pas \^etre lu.

C'est un gros travail que vous vous proposez d'entreprendre, et je vous
adresse d'avance tous mes remerciements.

Bien cordialement,

Michel H\'enon}
  \end{quotation}

\section*{Acknowledgements}
I am indebted to Florent Renaud for much of the material in Section \ref{sec:epilogue}.

\label{lastpage}

\end{document}